\documentclass[prl,floatfix,amsfonts]{revtex4}
\usepackage{graphicx,graphics,color,epsfig}
\usepackage{bm}
\usepackage{amsmath}
\usepackage{amssymb}

\newcommand{\bk}{{\bf k}}

\begin{document}
\title{Erratum: Nodal Cooper-pair stabilized phase dynamics in granular 
$d$-wave superconductors}
\author{ Yogesh N. Joglekar, A. H. Castro Neto, and Alexander V. Balatsky}
\maketitle


Recently, we discovered a mistake in our Letter~\cite{jj} on phase dynamics 
in granular $d$-wave superconductors, which we corrected in an Erratum 
published last week~\cite{jjer}. This is an expanded version of the Erratum, 
containing details and supporting numerical results.  

In our Letter, we accounted for weak disorder in $d$-wave superconducting 
state by adding lifetimes to the nodal quasiparticles  
$E_\bk\rightarrow E_\bk-i\gamma$ ($\gamma>0$) {\it when evaluating thermal 
Green's functions} $G(\bk,\tau)$ and $F(\bk,\tau)$. This naive substitution 
in thermal Green's functions leads to $G_{dis}\sim G\exp (i\gamma\tau)$ and 
$F_{dis}\sim F\exp (i\gamma\tau)$. However, the exact expressions for thermal 
Green's functions at temperature $T$ are given by 
\begin{eqnarray}
\label{eq:GT}
G(\bk,\tau)& = & -2T\sum_{n=0}^{\infty}\frac{(\omega_n+\gamma)
\sin\omega_n\tau}{(\omega_n+\gamma)^2+E_\bk^2},\\
\label{eq:FT}
F(\bk,\tau) & = & -2T\sum_{n=0}^{\infty}\frac{\Delta_\bk\cos\omega_n\tau}
{(\omega_n+\gamma)^2+E_\bk^2},
\end{eqnarray}
where $E_\bk^2=(\epsilon_\bk-\mu)^2+\Delta_\bk^2$ is the quasiparticle 
dispersion in the superconducting state, $\Delta_\bk=\Delta_0\cos2\theta_\bk$ 
is the $d$-wave superconducting gap, $\omega_n=T\pi(n+1/2)$ are fermionic 
Matsubara frequencies, and we have characterized the self-energy due to 
disorder as $\Im\Sigma(i\omega_n)=-\gamma\mathrm{ sign}(\omega_n)$.

The thermal Green's functions, Eqs.(\ref{eq:GT}) and (\ref{eq:FT}), are 
purely real and satisfy the following relations
\begin{eqnarray}
\label{eq:Gprop}
G(\bk,\tau) & = & -G(\bk,-\tau)=+G(\bk,1/T-\tau),\\ 
\label{eq:Fprop}
F(\bk,\tau) & = & +F(\bk,-\tau)=-F(\bk,1/T-\tau),
\end{eqnarray}
where $-1/T\leq\tau\leq 1/T$. Consequently, they also satisfy 
$G(1/2T-\tau)=G(1/2T+\tau)$ and $F(1/2T-\tau)=-F(1/2T+\tau)$ where 
$|\tau|\leq 1/2T$. The Matsubara frequency sums above can be 
re-expressed in 
terms of sum-over-residues, and an integral over a branch cut which cannot be 
performed analytically. In the $T=0$ limit, the sums reduce to 
\begin{eqnarray}
\label{eq:G}
G(\bk,\tau)& = & -\int^{\infty}_0\frac{d\omega}{2\pi}\frac{2(\omega+\gamma)
\sin\omega\tau}{(\omega+\gamma)^2+E^2_\bk},\\
\label{eq:F}
F(\bk,\tau)& = & -\int^{\infty}_0\frac{d\omega}{2\pi}\frac{2\Delta_\bk
\cos\omega\tau}{(\omega+\gamma)^2+E^2_\bk}.
\end{eqnarray}
These integrals can be expressed in terms of Sine ($I_s$) and Cosine ($I_c$) 
integrals~\cite{gr}
\begin{eqnarray}
\label{eq:GE}
G(\bk,\tau)& = & -\frac{1}{2\pi}\left[I_s(\tau,\gamma-iE_\bk)+I_s(\tau,\gamma+
iE_\bk)\right],\\
\label{eq:FE}
F(\bk,\tau)& = & -\frac{\Delta_\bk}{2\pi iE_\bk}\left[I_c(\tau,\gamma-iE_\bk)
-I_c(\tau,\gamma+iE_\bk)\right].
\end{eqnarray}
but the results, Eqs.(\ref{eq:GE}) and (\ref{eq:FE}), are not particularly 
illuminating. They can be expressed in terms of elementary functions 
only in the clean limit $\gamma\rightarrow 0$, in which case they reduce to 
standard analytical results. 

\begin{figure}[h]
\begin{center}
\begin{minipage}{20cm}
\begin{minipage}{9cm}
\epsfxsize=3in
\epsffile{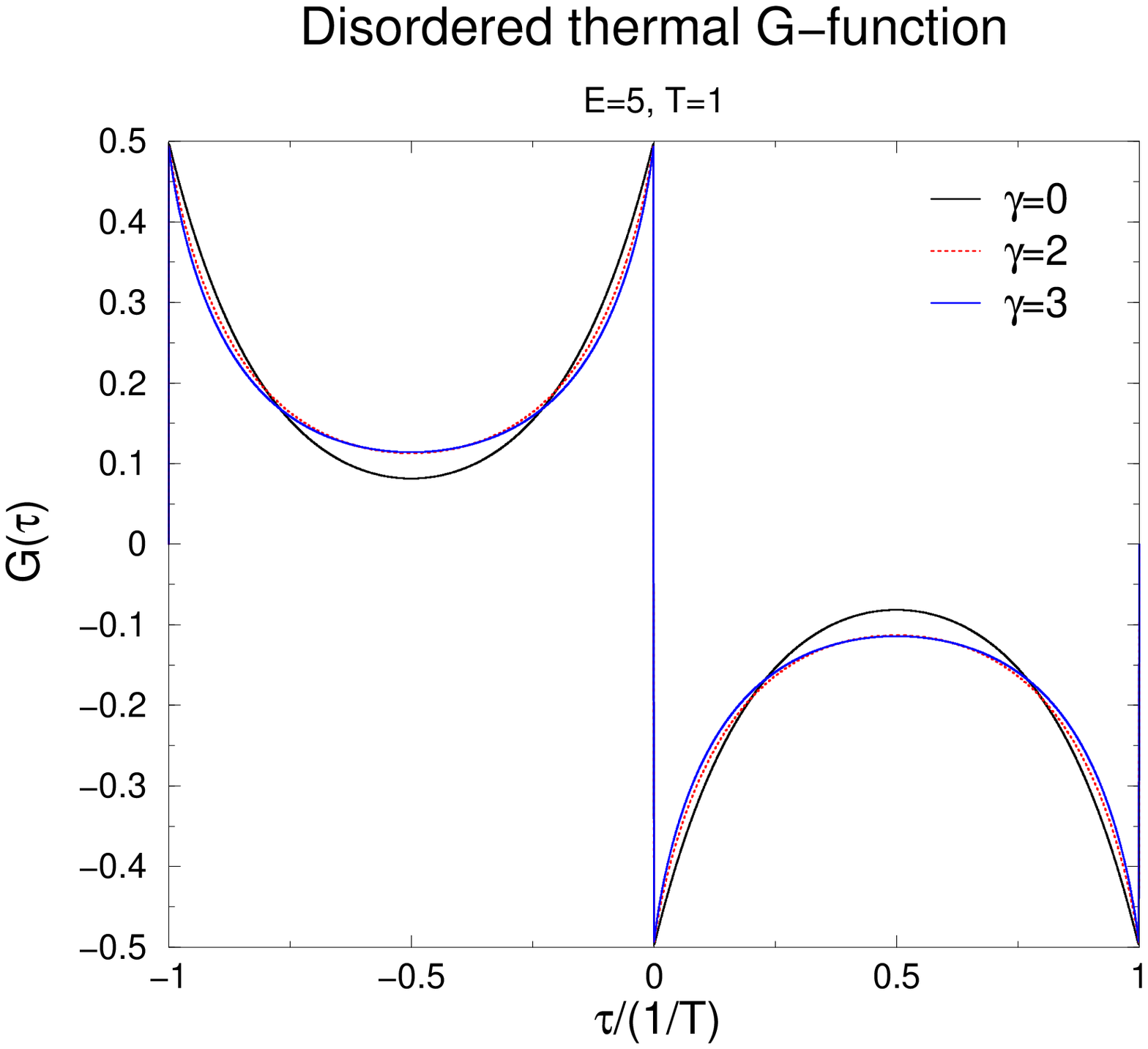}
\end{minipage}
\begin{minipage}{9cm}
\epsfxsize=3in
\epsffile{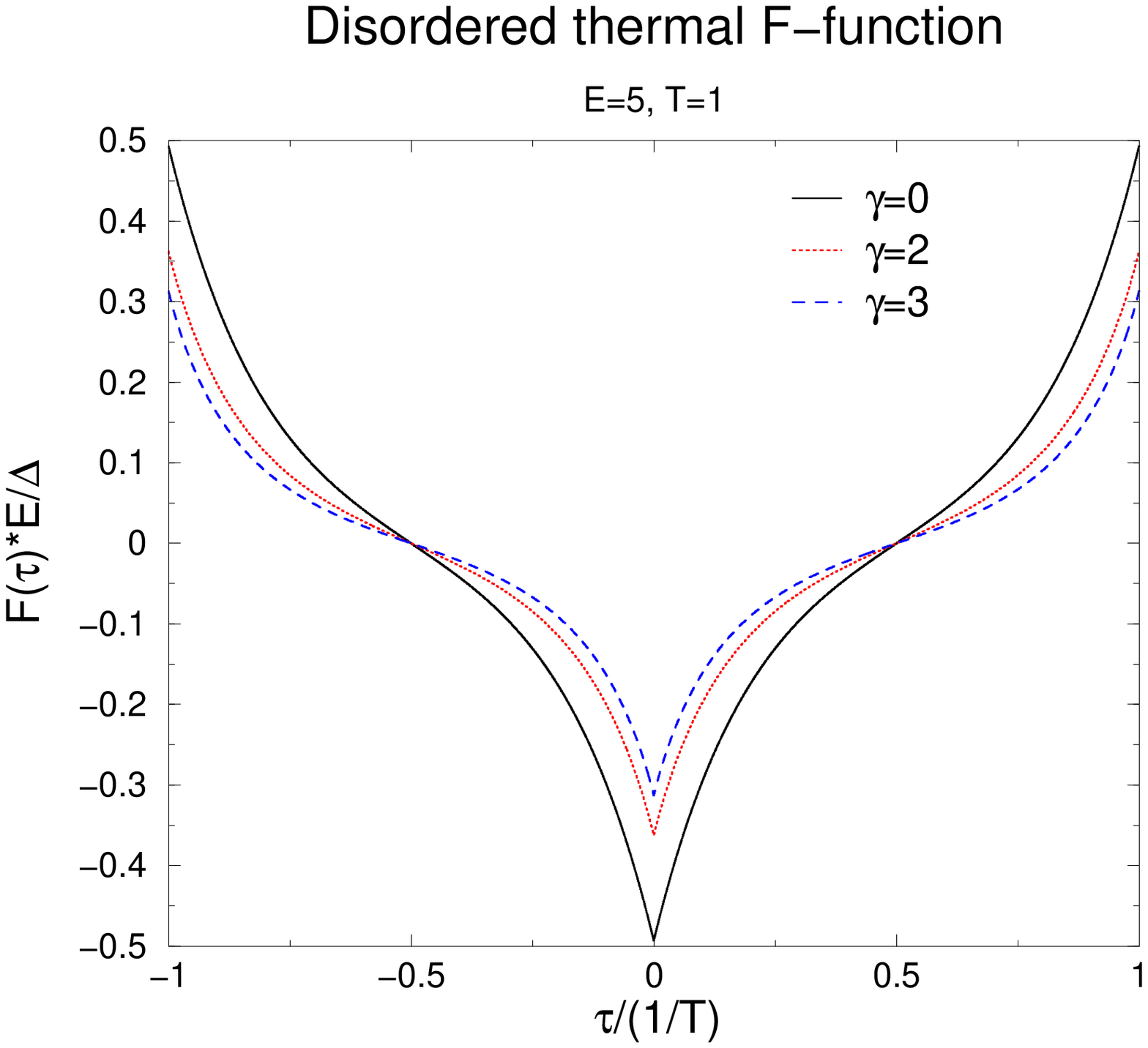}
\end{minipage}
\end{minipage}
\vspace{-0.5cm}
\caption{Normal and anamolous propagators, $G(\tau)$ and $F(\tau)$, 
calculated numerically for different disorder strengths $\gamma$ by using Eqs.
(\ref{eq:GT}) and (\ref{eq:FT}). The disordered Green's functions satisfy 
the symmetry properties in Eqs.(\ref{eq:Gprop}) and (\ref{eq:Fprop}).}
\label{fig:FGall}
\end{center}
\end{figure}

\begin{figure}[htbp]
\begin{center}
\begin{minipage}{20cm}
\begin{minipage}{9cm}
\epsfxsize=3in
\epsffile{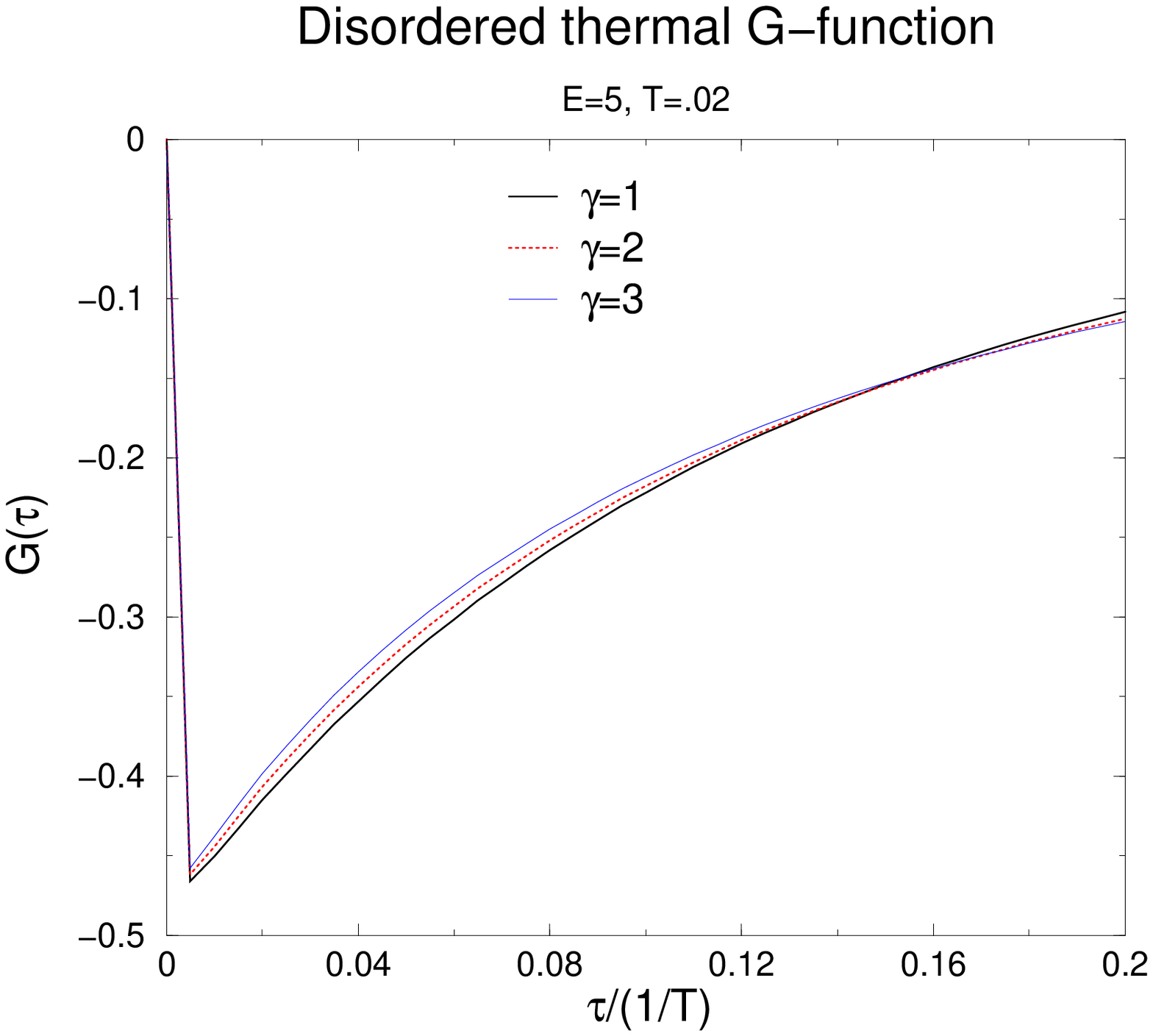}
\end{minipage}
\begin{minipage}{9cm}
\epsfxsize=3in
\epsffile{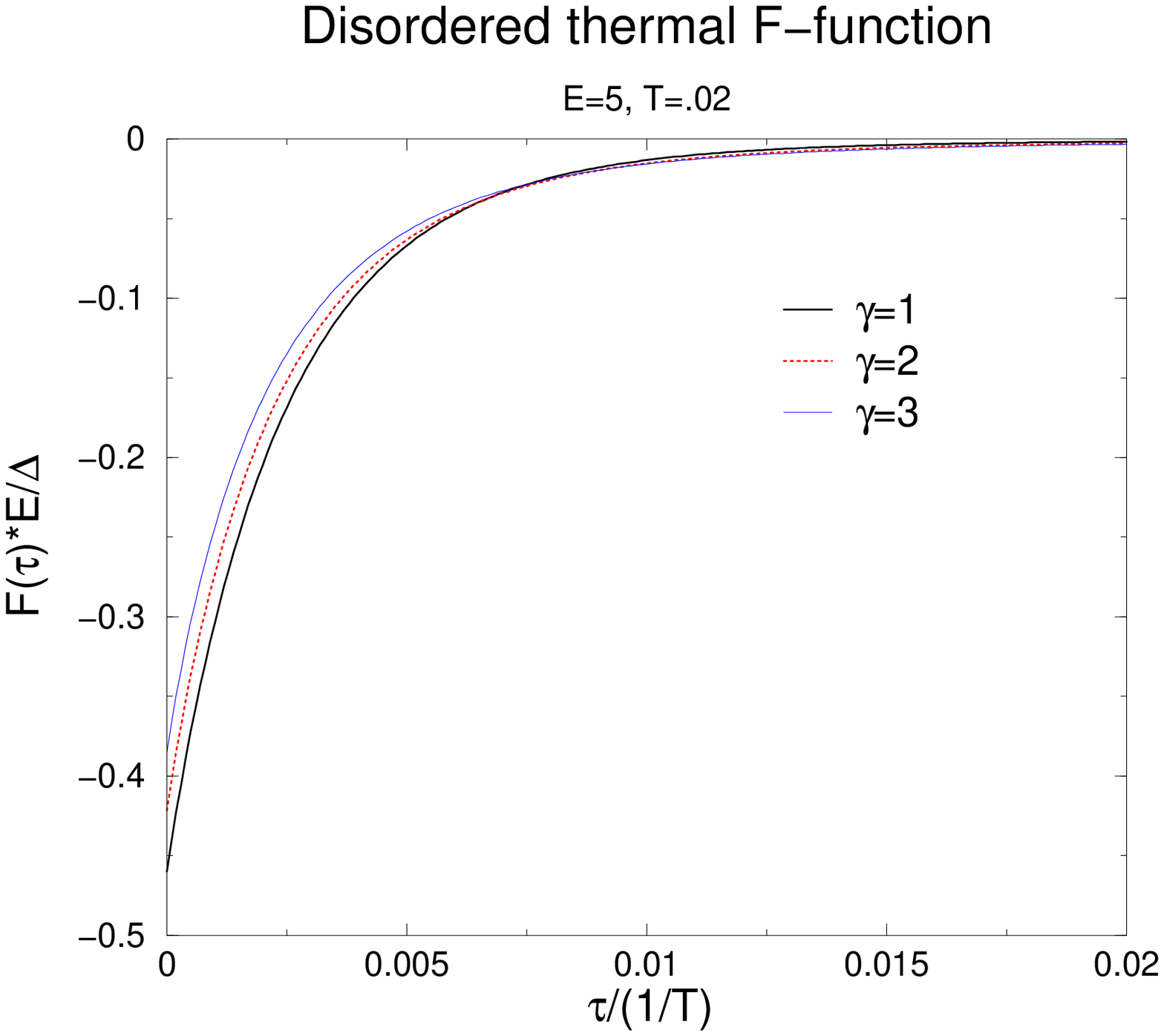}
\end{minipage}
\end{minipage}
\caption{$G$ and $F$ functions for $T\ll E,\gamma$. It is clear that both 
decay with increasing disorder. The kink in $G(\tau)$ at small $\tau$ 
is due to the discontinuity in the single-fermion propagator at $\tau=0$.}
\label{fig:FGlong}
\end{center}
\end{figure}

\begin{figure}[htbp]
\begin{center}
\epsfxsize=3in
\epsffile{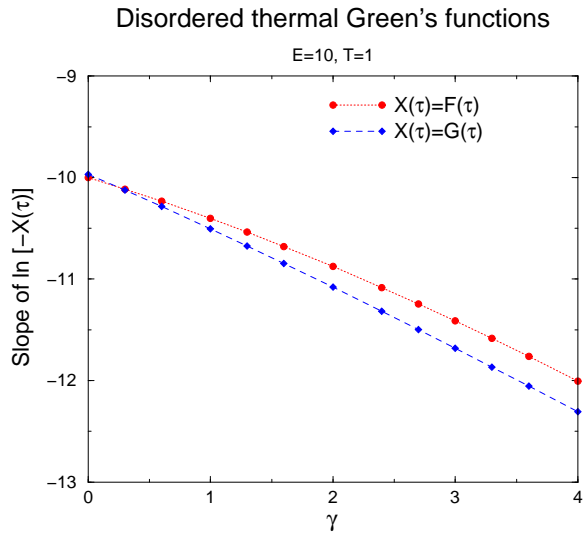}
\caption{Dependence of $G(\tau)$ and $F(\tau)$ on disorder, characterized 
by using the fit $G_{dis}=G\exp(-a_G\gamma|\tau|)$ and 
$F_{dis}=F\exp(-a_F\gamma|\tau|)$, for $E=10$. In the present case, 
$a_G=0.6$ and $a_F=0.5$. The intercept on the $y$-axis is $-E=-10$ and is 
consistent with the analytical result for low temperatures.}
\label{fig:FGslope}
\end{center}
\end{figure}

It is straightforward to evaluate the frequency sums in Eqs.(\ref{eq:GT}) and 
(\ref{eq:FT}) numerically. Figure~\ref{fig:FGall} shows $G(\tau)$ and 
$F(\tau)$ for $-1/T\leq\tau\leq 1/T$ for different 
disorder strengths. The Reader can see that the numerical results obey all 
symmetry properties listed in the last paragraph. At low temperatures, the 
long-time behavior of the disordered thermal Green's functions is shown in 
Fig.\ref{fig:FGlong}. It is clear that, for $T\ll \Delta_0,\gamma$ and 
$\tau T\ll 1$, both 
$G(\tau)$ and $F(\tau)$ decay with disorder. We can approximate them as 
$G_{dis}(\tau)=G(\tau)\exp(-a_G\gamma|\tau|)$ and 
$F_{dis}(\tau)=F(\tau)\exp(-a_F\gamma|\tau|)$, where $a_G,a_F\sim O(1)$ (See 
Fig.~\ref{fig:FGslope}). 

Therefore, although Eqs.(9) and (12) in~\cite{jj} are correct, our results 
$\alpha_{dis}(\tau)=\cos(2\gamma\tau)\alpha(\tau)$, 
$\beta_{dis}(\tau)=\beta(\tau)$ are incorrect. 
Instead, we find that the disorder suppresses the power-law correlations in 
Eqs.(9) and (12) in\cite{jj}, 
\begin{eqnarray}
\alpha_{dis}(\tau)& \sim (G_LG_R+G^{*}_RG^{*}_L)= & 
e^{-2a_G\gamma|\tau|}\alpha(\tau),\\
\beta_{dis}(\tau)& \sim (F_LF^{*}_R+F_RF^{*}_L)= & 
e^{-2a_F\gamma|\tau|}\beta(\tau).
\end{eqnarray}
Since the disorder suppresses the long-range phase correlations, it is 
necessary to exercise caution in using gradient expansion to map the 
nonlocal imaginary-time phase action (Eq.(4) in~\cite{jj}) on to a 
local Lagrangian for a Josephson junction (Eq.(17) in~\cite{jj}). 
In particular, the (apparently) divergent contribution to capacitance, 
which arises from momentum-conserving tunneling term (Eq.(16) in~\cite{jj}) 
is incorrect. Instead the junction capacitance is determined by the 
point-contact 
tunneling term, $C\sim A_0\Delta_0\sim |T_0|^2/\Delta_0$, and the 
Josephson energy is determined by the momentum-conserving tunneling term. 
Subsequently, our prediction that the Josephson plasmon frequency is strongly 
renormalized due to long-range phase correlations and is not sensitive to 
disorder, $\omega_p\sim\sqrt{\Delta_0T}$, Eq.(20) in~\cite{jj} is not valid. 
The correct dispersion is $\omega_p\sim\Delta_0$ and it is weakly renormalized 
due to disorder.

Thus, our results on robustness of $d$-wave Josephson junctions against 
disorder and the strong renormalization of the Josephson plasmon frequency 
are invalid. The clean-limit results in our paper, such as power-law 
phase-correlations and the subsequent sub-ohmic dissipation, continue to 
remain valid. 

We wish to thank Professor Dmitri Khveshchenko for his questions which led to 
the discovery of our mistake. 

\end{document}